 \documentclass[reprint, amsmath, amssymb, aps, prb]{revtex4-1}

\usepackage[]{graphicx}      
\usepackage{bm}             	
\usepackage{times}			
\usepackage{tabularx}
\usepackage{color}
\usepackage{dcolumn}		
\usepackage{amsmath}
\usepackage{amssymb}
\usepackage{amsfonts}
\usepackage{times}
\usepackage{tabularx}
\usepackage{xcolor,colortbl}

\begin{document}
\title{Using rf voltage induced ferromagnetic resonance to study the spin-wave density of states and the Gilbert damping in perpendicularly magnetized disks}
\author{Thibaut Devolder}
\email{thibaut.devolder@u-psud.fr}
\affiliation{Centre de Nanosciences et de Nanotechnologies, CNRS, Univ. Paris-Sud, Universit\'e Paris-Saclay, C2N-Orsay, 91405 Orsay cedex, France}

\date{\today}                                           
%
%
\begin{abstract}
We study how the shape of the spinwave resonance lines in rf-voltage induced FMR can be used to extract the spin-wave density of states and the Gilbert damping within the precessing layer in nanoscale magnetic tunnel junctions that possess perpendicular magnetic anisotropy. We work with a field applied along the easy axis to preserve the cylindrical symmetry of the uniaxial perpendicularly magnetized systems. We first describe the experimental set-up to study the susceptibility contributions of the spin waves in the field-frequency space. We then identify experimentally the maximum device size above which the spinwaves confined in the free layer can no longer be studied in isolation as the linewidths of their discrete responses make them overlap into a continuous density of states. The rf-voltage induced signal is the sum of two voltages that have comparable magnitudes: a first voltage that originates from the linear transverse susceptibility and rectification by magneto-resistance and a second voltage that arises from the non-linear longitudinal susceptibility and the resultant time-averaged change of the exact micromagnetic configuration of the precessing layer. The transverse and longitudinal susceptibility signals have different dc bias dependences such that they can be separated by measuring how the device rectifies the rf voltage at different dc bias voltages. The transverse and longitudinal susceptibility signals have different lineshapes; their joint studies in both fixed field-variable frequency, or fixed frequency-variable field configurations can yield the Gilbert damping of the free layer of the device with a degree of confidence that compares well with standard ferromagnetic resonance. Our method is illustrated on FeCoB-based free layers in which the individual spin-waves can be sufficiently resolved only for disk diameters below 200 nm. The resonance line shapes on devices with 90 nm diameters are consistent with a Gilbert damping of $0.011$. \textcolor{black}{A single value of the damping factor accounts for the line shape of all the spin-waves that can be characterized}. This damping of 0.011 exceeds the value of 0.008 measured on the unpatterned films, which indicates that device-level measurements are needed for a correct evaluation of dissipation. 
\end{abstract}

\maketitle

%
%



The frequencies of the magnetization eigenmodes of magnetic body reflect the energetics of the magnetization. As a result the frequency-based methods -- the ferromagnetic resonances (FMR)\cite{kittel_theory_1948} and more generally the spin-wave spectroscopies-- are particularly well designed for the metrology of the various magnetic interactions. In particular, measuring the Gilbert damping parameter $\alpha$ that describes the coupling of the magnetization dynamics to the thermal bath, specifically requires high frequency measurements. 
There are two main variants of these resonance techniques. The so-called conventional FMR and its modern version the vector network analyzer\cite{bilzer_vector_2007} (VNA)-FMR are established technique to harness the coupling of microwave photons to the magnetization eigenmodes to measure to anisotropy fields\cite{kittel_theory_1948}, demagnetizing fields, exchange stiffness\cite{kubota_structure_2009}, interlayer exchange\cite{devolder_ferromagnetic_2016} and spin-pumping\cite{mizukami_ferromagnetic_2001}, most often at film level. More recent methods, like the increasingly popular spin-transfer-torque-(STT)-FMR, are developed\cite{tulapurkar_spin-torque_2005}  to characterize the magnetization dynamics of magnetic bodies embodied in electrical devices possessing a magneto-resistance of some kind. 

In conventional FMR or VNA-FMR, the community is well aware that the line shape of a resonance is more complicated than simple arguments based on the Landau-Lifshitz-Gilbert equation would tell. There are for instance substantial contributions from microwave shielding effects \cite{bailleul_shielding_2013} ("Eddy currents") for conductive ferromagnetic films \cite{lin_rigorous_2015} or ferromagnetic films in contact with (or capacitively coupled to) a conductive layers. A hint to these effect is for instance to compare the lineshapes \cite{lin_rigorous_2015}  for the quasi-uniform precession mode and the first perpendicular standing spin wave modes that occur in different resonance conditions. Note that the experimental lineshapes are already complex in VNA-FMR despite the fact that the dynamics is induced by simple magnetic fields supposedly well controlled.

In contrast, STT-FMR methods rely on torques [spin-orbit torques (SOT)\cite{liu_spin-torque_2011} or STT] that have less hindsight that magnetic fields or that are the targeted measurements. These torques are related to the current across the device and the experimental analysis generally assumes that this current is \textit{in phase} with the applied voltage. This implicitly assumes that the sample is free of capacitive and inductive responses, even at the microwave frequencies used for the measurement. A careful analysis is thus needed when the STT-FMR methods analyze the phase of the device response to separate the contribution of the different torques\cite{tulapurkar_spin-torque_2005, sankey_measurement_2008, kubota_quantitative_2008}. Besides, the quasi-uniform mode is often the sole to be analyzed despite that fact that the line shapes of the higher frequency modes can be very different \cite{sankey_measurement_2008}. Finally, an external field is generally applied in a direction that is not a principal direction of the magnetization energy functional \cite{cheng_nonlinear_2013}. While this maximizes the signal, this unfortunately makes numerical simulation unavoidable to model the experimental responses. 

With the progress in MTJ technologies, much larger magneto-resistance are now available\cite{ikeda_tunnel_2008}, such that signals can be measured while maintaining sample symmetries, for instance with a static field applied collinearly to the magnetization. In addition, high anisotropy materials can now be incorporated in these MTJs. This leads to a priori much more uniform magnetic configurations in which analytical descriptions are more likely to apply. In this paper, we revisit \textit{rf}-voltage induced FMR in a situation where the symmetry is chosen so that all torques should yield a priori the same canonical lineshape for all spinwaves excited in the system. We use PMA MTJ disks of sizes 500 nm, on which a quasi-continuum of more that 20 different spin-wave modes can be detected, down to sizes of 60 nm where only a few discrete spinwave modes can be detected. We discuss the lineshapes of the spin-wave signals with the modest objective of determining if at least the Gilbert damping of the dynamically active magnetic layer can be reliably extracted. We show that the linear transverse susceptibility and the non-linear longitudinal susceptibilities must both be considered when a finite \textit{dc} voltage is applied through the device. We propose a methodology and implement it on a nanopillars made with a standard MgO/FeCoB/MgO free layer system in which we obtain a Gilbert damping of $0.011\pm 0.0003$. This exceeds the value of 0.008 measured on the unpatterned film, which indicates that device-level measurements are needed for a correct evaluation of dissipation.

The paper is organized as follows: \\
The first section lists the experimental considerations, including the main properties of the sample, the measurement set-up and the mathematical post-processing required for an increased sensitivity. The second section discusses the origins of the measured resonance signals and their main properties. The third section describes how the device diameter affects the spin-wave signals in \textit{rf}-voltage-induced ferromagnetic resonance. The last section describes how the voltage bias dependence of the spinwave resonance signals can be manipulated to extract the Gilbert damping of the dynamically active magnetic layer. After the conclusion, an appendix details the main features of the spectral shapes expected in ideal perpendicularly magnetized systems.

\section{Experimental considerations}

\subsection{Magnetic tunnel junctions samples}
We implement our characterization technique on the samples described in detail in ref.~\onlinecite{devolder_size_2016}. They are tunnel junctions with an FeCoB-based free layer and a hard reference system based on a well compensated \textcolor{black}{[Co/Pt]-based} synthetic antiferromagnet. All layers have perpendicular magnetic anisotropy (PMA). \textcolor{black}{The perpendicular anisotropy of the thick ($t=2~\textrm{nm}$) free layer is ensured by a dual MgO encapsulation and an iron-rich composition. After annealing, the free layer has an areal moment of $M_s t \approx 1.8~\textrm{mA}$ and an effective perpendicular anisotropy field $\mu_0(H_k-M_s)$ = 330~\textrm{mT}. Before pattering, standard ferromagnetic resonance measurements indicated a Gilbert damping parameter of the free layer being $\alpha = 0.008$}. Depending on the size of the patterned device, the tunnel magnetoresistance (TMR) is 220 to 250\%, for a stack resistance-area product is $\textrm{RA}=12~ \Omega.\mu \textrm{m}^2$. The devices are circular pillars with diameters varied from 60 to 500 nm. The materials, processing and device \textit{rf} circuitry were optimized for fast switching \cite{devolder_size_2016} spin-transfer-torque magnetic random access memories (STT-MRAM\cite{khvalkovskiy_basic_2013}) ; the quasi-static \textit{dc} switching voltage is $\approx 600$ mV. In the present report, the applied voltages shall never exceed 100 mV to minimize spin-transfer-torque effects. The fields will always be applied along (z) which is the easy magnetization axis. The sample will be maintained in the antiparallel (AP) state.

\subsection{Measurement set-up}
\begin{figure}
\includegraphics[width=9 cm]{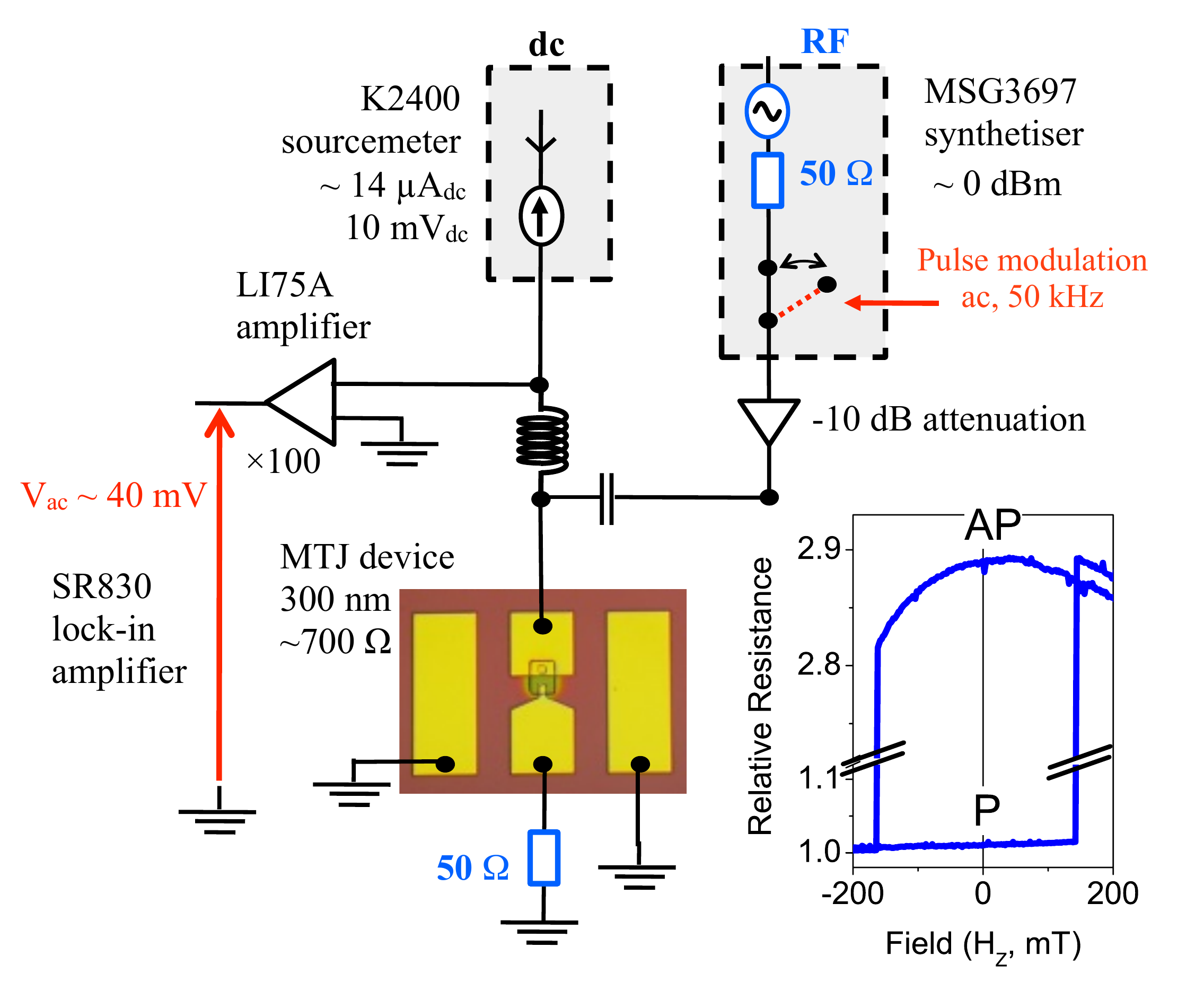}
\caption{(Color online). Sketch of the experimental set-up with an $300 \times 300~\mu \textrm{m}^2$ optical micrograph of the device circuitry. The given numbers are the typical experimental parameters for a 300 nm diameter junction. Inset: resistance versus out-of-plane field hysteresis loop for a device with 300 nm diameter.}
\label{T}
\end{figure}

The pillars are characterized in a set-up (Fig.~\ref{T}) inspired from spin-torque diode experiments \cite{tulapurkar_spin-torque_2005} but an electrical bandwidth increased to 70 GHz. The objective is to identify the regions in the \{frequency, field\} space in which the magnetization is responding in a resonant manner. The device is attacked with an \textit{rf} voltage $V_{rf}$. A 10 dB attenuator is inserted at the output port of the synthesizer to improve its impedance matching so as to avoid standing waves in the circuit. This improves the frequency flatness of the amplitude of the stimulus arriving at the device. To ease the detection of the sample's response, the \textit{rf} voltage is pulse-modulated at an \textit{ac} frequency $\omega_{ac} / (2\pi) = 50~\textrm{kHz}$ (Fig.~\ref{T}). The current passing through the MTJ has thus frequency components at the two sidebands $\omega_{rf} \pm \omega_{ac}$. The \textit{ac} voltage which appears across the device is amplified and analyzed by a lock-in amplifier. We shall discuss the origin of this \textit{ac} voltage in section~\ref{demodulation}. Optionally, the device is biased using a \textit{dc} sourcemeter supplying $V_\textrm{dc}$ and measuring $I_{dc}$. 

%

\begin{figure}
\includegraphics[width=8.5 cm]{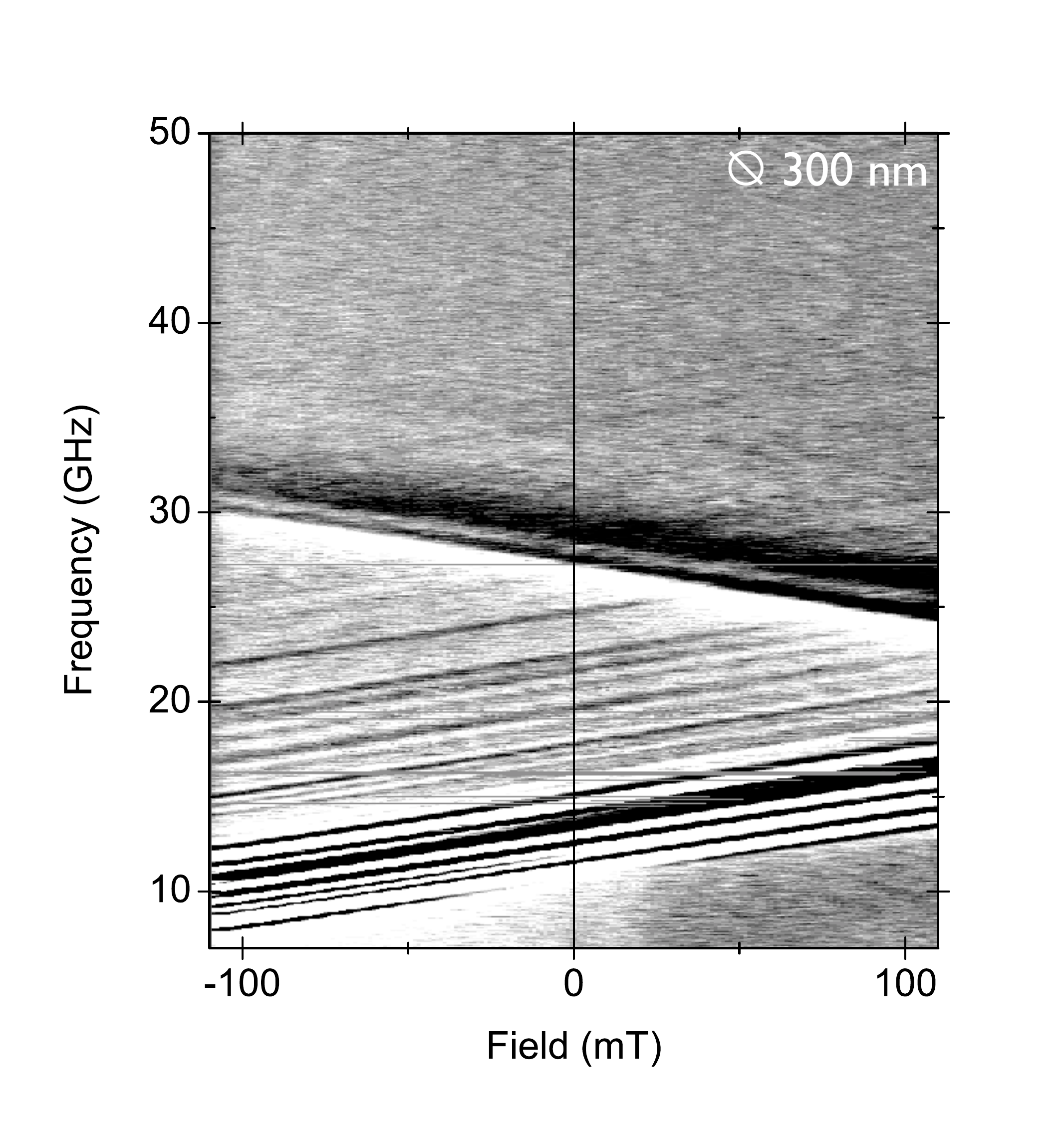}
\caption{Field derivative of the rectified voltage $\frac{dV_{ac}}{dH_z}$ in the \{frequency-field\} parameter space for a 300 nm diameter device in the AP state when the field is parallel to the free layer magnetization. The linear features with positive (resp. negative) slopes correspond to free layer (resp. reference layers) confined spin-wave modes. \textcolor{black}{Black and white colors correspond to signals exceeding $\pm 0.01~\textrm{V/T}$}. The one-pixel high horizontal segments are experimental artefacts due to transient changes of contact resistances.}
\label{map}
\end{figure}

Figure \ref{map} shows a representative map of the $\frac{dV_{ac}}{dH_z}$ response obtained on a pillar of diameter 300 nm with $V_{dc}=10~\textrm{mV}$. As positive fields are parallel to the free layer magnetization, the spin waves of the free layer appear with a positive frequency versus field slope, expected to be the gyromagnetic ratio $\gamma_0$ \textcolor{black}{of the free layer material} (see appendix). Conversely, the reference layer eigenmodes appear with a negative slope, expectedly $-\gamma_0$, \textcolor{black}{where this time $\gamma_0$ is gyromagnetic ratio of the reference layer material combination}. Working in the AP state is thus a convenient way to easily distinguish between the spinwaves of the free layer and of the reference layers. \textcolor{black}{Note that the gyromagnetic ratios $\gamma_0$ of the free layer mode and the reference layer modes differ slightly owing to their difference chemical nature. The free layer has a Land\'e factor $g=2.085\pm0.015$ where the error bar is given by the precision of the field calibration; the reference layer modes are consistent with a 1.2\% larger gyromagnetic ratio. The accuracy of this latter number is limited only by the signal-to-noise ratio in the measurement of the reference layer properties.}
 Looking at Fig. 2, one immediately notices that the linewidths of the reference layer modes are much broader than that of the free layer. While the linewidh of the reference layer modes will not be analyzed here, we mention that this increased linewidth is to be expected for reference layers that contain heavy metals \textcolor{black}{(Pt, Ru)} with large spin-orbit couplings, hence larger damping factors \cite{kambersky_spin-orbital_2007}.

%

\subsection{Experimental settings}  

In practice, we choose an applied field interval of $[-110, 110~\textrm{mT}]$ that is narrow enough to stay in \textcolor{black}{a state whose resistance is very close to that of the remanent AP state}. 
The frequency $\omega_{rf} / (2\pi)$ is varied from 1 to 70 GHz; we generally could not detect signals above 50 GHz. The practical frequency range $2\pi\times 50~\textrm{GHz} / \gamma_0 \approx 1.6~\textrm{T}$ is much wider that our accessible field range. For wider views of the experimental signals (for instance when the spin-wave density of states is the studied thing), we shall thus prefer to plot them versus frequency than versus field. 
The response is recorded pixel by pixel in in the \{frequency, field\} space. The typical pixel size is  $ \{ \delta H_z \times \delta f \}=$ \{1 mT $\times$ 50 MHz\}. The field and frequency resolutions are thus comparable (indeed $2\pi\times \delta f / \gamma_0 = 1.7$ mT). 
\subsection{Signal conditioning} 
\subsubsection{Mathematical post-treatments} \label{post}
Finally, despite all our precautions to suppress the rectifying phenomena that do not originate from magnetization dynamics, we have to artificially suppress the remaining ones. This was done by mathematical differentiation, and we generally plot $\frac{dV_{ac}}{df}$ or $\frac{dV_{ac}}{dH_z}$ in the experimental figures (Figs.~\ref{map}-\ref{BiasDep}).

\subsubsection{Dynamic range improvement by self-conformal averaging} \label{conformal}
\begin{figure}
\includegraphics[width=9 cm]{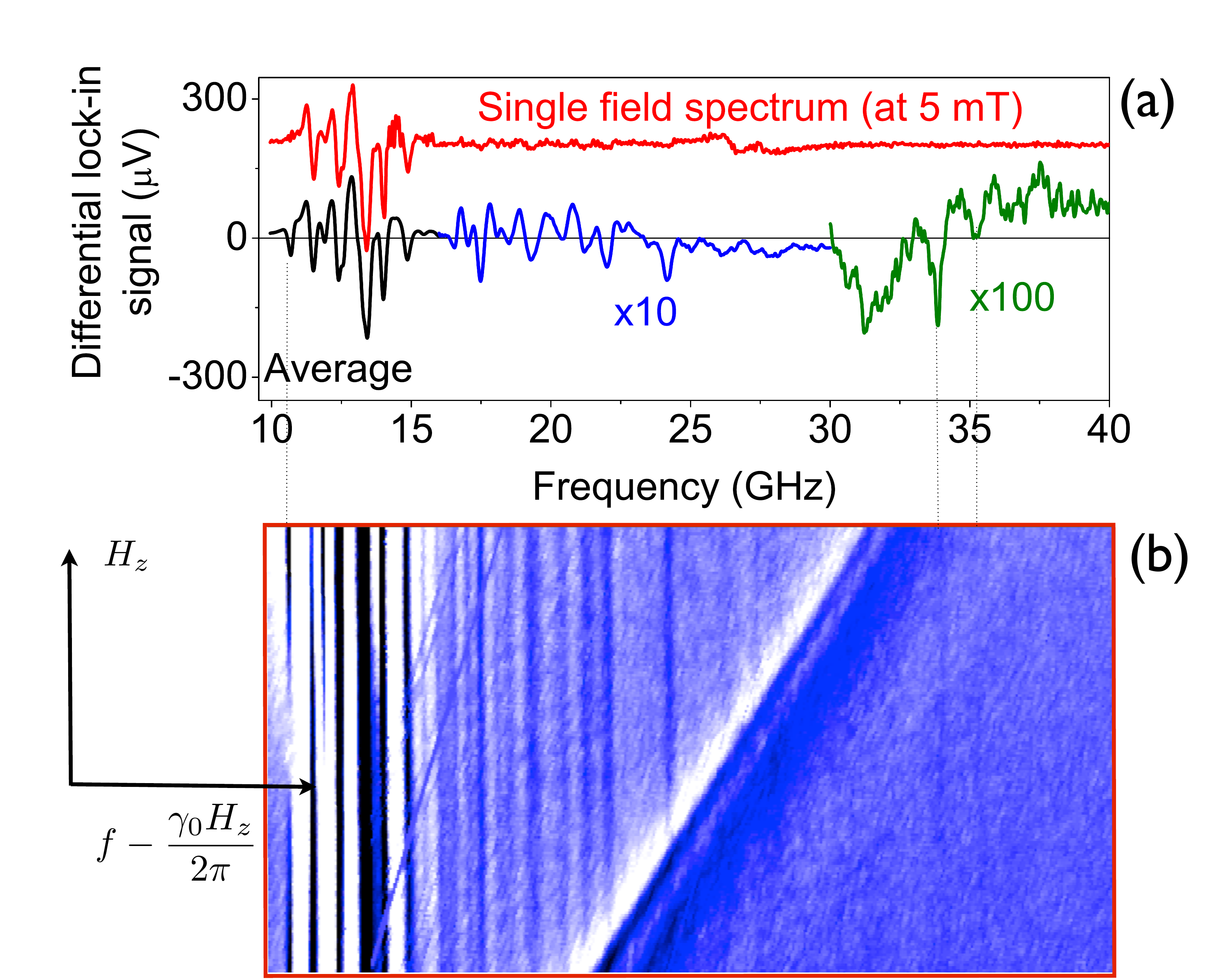}
\caption{(Color online). Illustration of the dynamic range improvement by self-conformal averaging (section \ref{conformal}). The procedure is implemented on a 300 nm diameter device to evidence the free layer modes. Bottom panel: field derivative of the \textit{ac} signal in the rotated frame in which the modes with $\frac{df}{dH_z} = \frac{\gamma_0}{2\pi}$ should appear as vertical lines. Top panel: comparison of a single field frequency scan (red) with the average over all scans as performed in the $\omega = \gamma_0 H_z$ direction. Note that the signal of the lowest frequency mode (which corresponds to the quasi-uniform precession) disappears near zero field, at 5 mT (see the apparent break in the middle of the most left line in the bottom panel).}
\label{slice}
\end{figure}

A special procedure (Fig.~\ref{slice}) is applied when a better signal to noise ratio is desired while the exact signal lineshape and amplitudes are not to meant to be looked at. This procedure harnesses the fact that the normalized shape of the sample's response is essentially self-conformal when moving across a line with $\frac{d\omega}{dH_z} = \gamma_0$ in the \{frequency, field\} parameter space (see appendix). 
The procedure consists in calculating the following primitive:

\begin{equation}
s(f_0) =  \frac{1}{2\gamma_0 H_z^{max}}\int_\textrm{contour} \frac{dV_\textrm{ac}}{dH_z} df~Ê,
   \label{test}
\end{equation}
in which the integration contour is the segment linking the points ($- H_z^{max}, f_0 - \gamma_0 H_z^{max}$) and ($ H_z^{max}, f_0 + \gamma_0 H_z^{max}$) in the \{field, frequency\} parameter space. Such contours appear as pixel columns in Fig.~\ref{slice}(b). 
This primitive (eq.~\ref{test}) is efficient to reveal the free layer spin-wave modes that yield an otherwise too small signal. For instance when only 7 modes can be detected in single field spectra [Fig.~\ref{slice}(a)], the averaging procedure can increase this number to typically above 25.
The averaging procedure is also effective in suppressing the signals of the reference layer as these laters average out over a contour designed for the free layer mode when in the AP state. However as the linewidth of the free layer modes is proportional to the frequency, it is not constant across the contour; the higher signal to noise ratio is thus unfortunately obtained at the expense of a distorted (and unphysical) lineshape. Note also that this procedure can not be applied to the quasi-uniform precession mode as will be explained in section \ref{qup}).

%

\section{Origin and nature of the rectified signal} \label{demodulation}
Let us now discuss the origin of the demodulated \textit{ac} voltage. In this section, we assume that the reference layer magnetization is static but  not necessarily uniformly magnetized. We can thus express any change of the resistance by writing $\delta R = \frac{\delta R}{\delta M} \delta M$ where $\delta$ has to be understood as a functional derivative with respect to the free layer magnetization distribution.

\subsection{The two origins of the rectified signals}
The \textit{ac} signal can contain two components $V_{1,~ac}$  and $V_{2,~ac}$ of different physical origins \cite{devolder_exchange_2016}. The first component is the 'standard' STT-FMR signal: the pulse-modulated \textit{rf} current is at the frequency sidebands $\omega_{rf} \pm \omega_{ac}$ and it rectifies to \textit{ac} any oscillation of the resistance $\delta R_{rf} $ occurring at the frequency $\omega_{rf}$. We simply have $V_{1,~ac} = \delta R_{{rf}} \times i_{\omega_{rf} \pm \omega_{ac}}$.

The second \textit{ac} signal ($V_{2,~ac}$) is related to the change of the time-averaged resistance due to the population of spinwaves created when the \textit{rf} current is applied \cite{cheng_nonlinear_2013}. Indeed the time-averaged magnetization distribution is not the same when the \textit{rf} is \textit{on} or \textit{off}. This change of resistance $\delta R_{ac}$ can revealed by the (optional) \textit{dc} current $I_{dc}$ passing through the sample, i.e. $ V_{2, ~ac} =  \delta R_{ac} \times  I_{dc} $. 

\textcolor{black}{Note that a third rectification channel \cite{kupferschmidt_theory_2006} can be obtained by a combination of spin pumping and inverse spin Hall effect in in-plane magnetized systems \cite{mihalceanu_spin-pumping_2017}. This third rectification channel yields symmetric lorentzian lines when applied to PMA systems in out-of-plane applied fields (see eq. 23 in ref. \onlinecite{kupferschmidt_theory_2006}). Besides, the spin-pumping is known to be largely suppressed by the MgO tunnel barrier \cite{baker_spin_2016}, such that we will consider that we can neglect this third rectification channel from now on.}
In summary, we have:
\begin{equation}  V_{1,~ac} =  \frac{V_{rf}}{R + 50} ~\frac{\delta R}{\delta M} ~\delta M_{rf}   \textrm{~~~~and}
\label{V1} 
\end{equation}
\begin{equation}  V_{2, ~ac} =  \frac{V_{dc}}{R + 50} ~\frac{\delta R}{\delta M} ~\delta M_{ac} 
\label{V2} 
\end{equation}

This has important consequences.
 
\subsection{Compared signal amplitudes in the P and AP states} 
The first important consequence of Eq.~\ref{V1} and \ref{V2} is that the signal amplitude depends on the nature of the micromagnetic configuration. As intuitive, both $V_{1,~ac}$ and $V_{2,~ac}$ scale with how much the instantaneous device resistance depends on its instantaneous micromagnetic configuration. This is expressed by the \textit{sensitivity factor} $\frac{\delta R}{\delta M}$ which is essentially a magneto-resistance. 
We expect no signal when the resistance is insensitive to the magnetization distribution at first order (i.e. when $\frac{\delta R}{\delta M} \equiv 0$). 

In our samples, the shape of the hysteresis loop (Fig.~\ref{T}) seems to indicate that the free layer magnetization is very uniform when in the Parallel state. Consistently, the experimental rectified signal were found to be weak signals when in the P state.  Conversely, there is a pronounced curvature in the AP branch of the $R(H_z)$ hysteresis loop (see one example in the inset of Fig.~\ref{T}). This indicates that the resistance is much dependent on the exact magnetization configuration when in the AP state. Consistently, this larger $\frac{\delta R}{\delta M}$ in the AP state is probably the reason why the rectified signal is much easier to detect in the AP state for our samples. 

\subsection{Bias dependence of the rectified signals} 
The second important consequence of Eqs.~\ref{V1}-\ref{V2} concerns the dependence of the rectified \textit{ac} signals $V_{1,~ac}$ and $V_{2,~ac}$ on the \textit{dc} and \textit{rf} stimuli. As $\delta M_{rf}$ scales with the applied \textit{rf} torque according to a linear transverse susceptibility ($\Re e(\chi_{xx})$, see appendix), $V_{1,~ac}$ is expected to scale with the \textit{rf} power ${V_{rf}}^2$ (see Eq.~\ref{V1}) independently from the \textit{dc} bias, i.e. we have $$V_{1, ~ac} \;\;\propto \;\;\ {V_{rf}}^2~.$$ 
In contrast, $\delta M_{ac}$ is related to a longitudinal susceptibility and is thus quadratic with the \textit{rf} torques (see appendix). Using Eq.~\ref{V2}, we thus expect the following bias dependence $$V_{2, ~ac} \;\;\propto \;\;\ {V_{rf}}^2 {V_{dc}}~. $$

\subsection{Peculiarities of the quasi-uniform precession (QUP) mode} 
The last important consequence of Eqs.~\ref{V1}-\ref{V2} concerns specifically the quasi-uniform precession (QUP) mode that shows a peculiar \textit{ac} signal. 
\subsubsection{Quasi-absence of STT-FMR like signal for the quasi-uniform precession mode}
In the idealized macrospin case (see appendix) the uniform precession is perfectly circular with no \textit{rf} variation of $M_z$ at any order. If the fixed layer was uniformly magnetized along exactly (z), this would lead $V_{1, ~ac} =0$ such that the signal of the QUP mode would be given by purely $V_{2, ~ac}$. This qualitatively 'pure $V_{2, ~ac}$ character' is confirmed experimentally by the fact that the signal of the QUP mode systematically changes sign with $V_{dc}$ in our sample series (not shown). 

\subsubsection{Strong dependence of the QUP signal amplitude with the applied field} \label{qup}
In addition, the experimental signal of the quasi-uniform mode is found to disappear at low fields [see Figs.~\ref{map}, \ref{slice}(b) and \ref{size}(a, b)] exactly at the apex of the AP branch of the hysteresis loop (Fig.~\ref{T}), i.e. for the field leading specifically to $\frac{dR}{dH_z}=0$. Moreover, the amplitude of the \textit{ac} signal of the QUP mode appears to be essentially linearly correlated with the loop slope $\frac{dR}{dH_z}$ (not shown). For instance, the $V_{2,~ac}$ of the QUP mode changes sign when the applied field crosses the apex of the $R(H_z)$ loops (Fig.~\ref{T}).

The reason stems probably from the sensitivity factor $\frac{\delta R}{\delta M}$ and its correlation with the loop slope $\frac{dR}{dH_z}$; in some sense, a large loop slope should translate in a large sensitivity factor. \textcolor{black}{While a numerical evaluation of this correlation goes beyond the scope of this paper, we stress that if the magnetization was perfectly uniform there would be a one-to-one correlation between loop slope $\frac{dR}{dH_z}$ and magnetoresistance sensitivity factor $\frac{\delta R}{\delta M}$. This trend remains qualitatively true for the QUP mode.} Indeed as the hysteresis loop is monitoring the \textit{spatial average} of the magnetization, it is more insightful for the \textit{uniform mode} than for any other (higher order) modes whose dynamic profiles spatially average to essentially zero \cite{klein_ferromagnetic_2008}; the correlation between $\frac{dR}{dH_z}$ and $\frac{\delta R}{\delta M}$ is thus expected to be maximal for quasi-uniform changes of the magnetization configuration. 

While this property -- the disappearance of the QUP mode signal when $\frac{dR}{dH_z}=0$ -- can be used to distinguish the QUP mode from the higher order spin waves, the pronounced field dependence of the QUP signal complicates the analysis, as it prevents to conveniently analyze the field derivative of the \textit{ac} signal (¤\ref{post}). In the remainder of this paper we shall focus on only higher order modes to avoid such difficulties.

\subsection{Signals for non-uniform spin-waves} 
Before analyzing the spin-wave density of states (section \ref{DOS}), let us comment on the amplitude of the STT-FMR-like signal  $V_{1,~ac}$ for the non-uniform spin-waves. In the perpendicular magnetization state, these spin-waves have a  circular precession \cite{kalinikos_theory_1986}. By symmetry, the resistance is not expected to change during a period of circular precession when in \textcolor{black}{the perfect collinear cases} \textcolor{black}{and for radial spin waves maintaining the cylindrical symmetry of the system}. In other words, when the dynamical magnetization of the eigenmode maintains the cylindrical symmetry and when the free and reference layers equilibrium magnetizations follow $\vec{M}_\textrm{free}  \times \vec{M}_\textrm{ref} = \vec{0}$ everywhere in the (xy) plane, \textcolor{black}{with $\times$ being the conventional vector product)} the device resistance is not expected to oscillate. While we can not identify to what extent we depart from this ideal situation, we speculate that this perfect collinearity does not happen in practice at least because of finite thermal fluctuations. \textcolor{black}{The effect of thermal fluctuations on the device resistance is not averaged out for non-uniform spin-waves, while it could be essentially averaged out for the QUP mode analyzed earlier}. In practice a finite variation of the resistance $\delta R_{rf}\neq 0$ is always present during a precession period for a non-uniform spin-wave. This provides a finite sensitivity to any spin-wave mode. This resistance variation at $\omega_{rf}$ has the spectral shape of a transverse susceptibility term $\Re e (\chi_{xx})$ (see appendix).

\begin{table*}
\begin{tabular}{lc|cccc|}
\hline
 \rule{0pt}{3ex}  $  $ $ $ $ $ $ $  Signal & & Spectral  &  Peak-to-peak  & Full Width & Zero crossings \\ 
 &    & shape &  separation & at Half Maximum & separation  \\ \hline
 \rule{0pt}{3ex} Expected signals and their stimulus dependence: & & \multicolumn{1}{c}{ } & \multicolumn{1}{c}{ } & \multicolumn{1}{c}{ } & \multicolumn{1}{c}{ } \vline \\
& & \multicolumn{1}{c}{ } & \multicolumn{1}{c}{ } & \multicolumn{1}{c}{ } & \multicolumn{1}{c}{ } \vline \\
 \rule{0pt}{0ex} $V_{1,~ac} \propto V_{rf}^2$  & & $\Re e(\chi_{xx})$   &  $2 \alpha \omega$ & - & -\\
 \rule{0pt}{0ex} $V_{2,~ac} \propto V_{rf}^2 V_{dc}$  & & $\Delta M_z$ &  - &  $2 \alpha \omega$ & - \\
& & \multicolumn{1}{c}{ } & \multicolumn{1}{c}{ } & \multicolumn{1}{c}{ } & \multicolumn{1}{c}{ } \vline \\ \hline
 \rule{0pt}{3ex} Signal extraction procedure from experiments: & & \multicolumn{1}{c}{ } & \multicolumn{1}{c}{ } & \multicolumn{1}{c}{ } & \multicolumn{1}{c}{ } \vline \\
& & \multicolumn{1}{c}{ } & \multicolumn{1}{c}{ } & \multicolumn{1}{c}{ } & \multicolumn{1}{c}{ } \vline \\
 \rule{0pt}{0ex} $\frac{d}{dH_z} V_{1,~ac}^{exp}$ estimated from $\frac{dV_{ac}^{exp}}{dH_z} \biggr \rvert_{V_{dc}=0}$  & & $\frac{d \Re e(\chi_{xx})}{d \omega}$ &  - & - & $2 \alpha \omega$  \\
 \rule{0pt}{7ex}   $\frac{d}{dH_z} V_{2,~ac}^{exp}$ estimated from $\left[ \frac{dV_{ac}^{exp}}{dH_z} \biggr \rvert_{V_{dc} \neq0} - \frac{dV_{ac}^{exp}}{dH_z} \biggr \rvert_{V_{dc}=0} \right]$ & & $\frac{d \Im m(\chi_{xx})}{d \omega}$ & $\frac{2}{\sqrt{3}} \alpha \omega$ &  - & - \\  
 &   & \multicolumn{1}{c}{ } & \multicolumn{1}{c}{ } & \multicolumn{1}{c}{ } & \multicolumn{1}{c}{ } \vline \\ \hline

\end{tabular}
\caption{Summary of the expected lineshapes and linewidths for the different signals that can be encountered in \textit{rf}-voltage-induced FMR experiments. An hyphen is inserted when the concept is not applicable.}

\label{PMAmacrospinLINESHAPES}
\end{table*}

\section{Spin wave density of states against lateral confinement} \label{DOS}
\textcolor{black}{Any reliable analysis of a spectral lineshape or linewidth requires to determine priorly how many spin-waves contribute to the lineshape under study. Therefore, before discussing the lineshapes of the individual spin-wave modes, let us determine how the lateral confinement influences the measured rectified signal. The impact of the device diameter on the spectral signals is reported in Fig.~\ref{size}.} 

\subsection{Spin-waves within the references layers}
Fig.~\ref{size} indicates that the modes of the reference layers have frequencies that are almost not affected by the device diameter. This fact is related to the well compensated character of the synthetic antiferromagnet that composes the reference layers. Indeed the internal demagnetizing fields compensate to some extent, such that they do not influence the frequency of the acoustical mode of a SAF as much as the anisotropies and the interlayer exchange couplings do.

\subsection{Spin-waves within the free layer}
\begin{figure}
\includegraphics[width=9 cm]{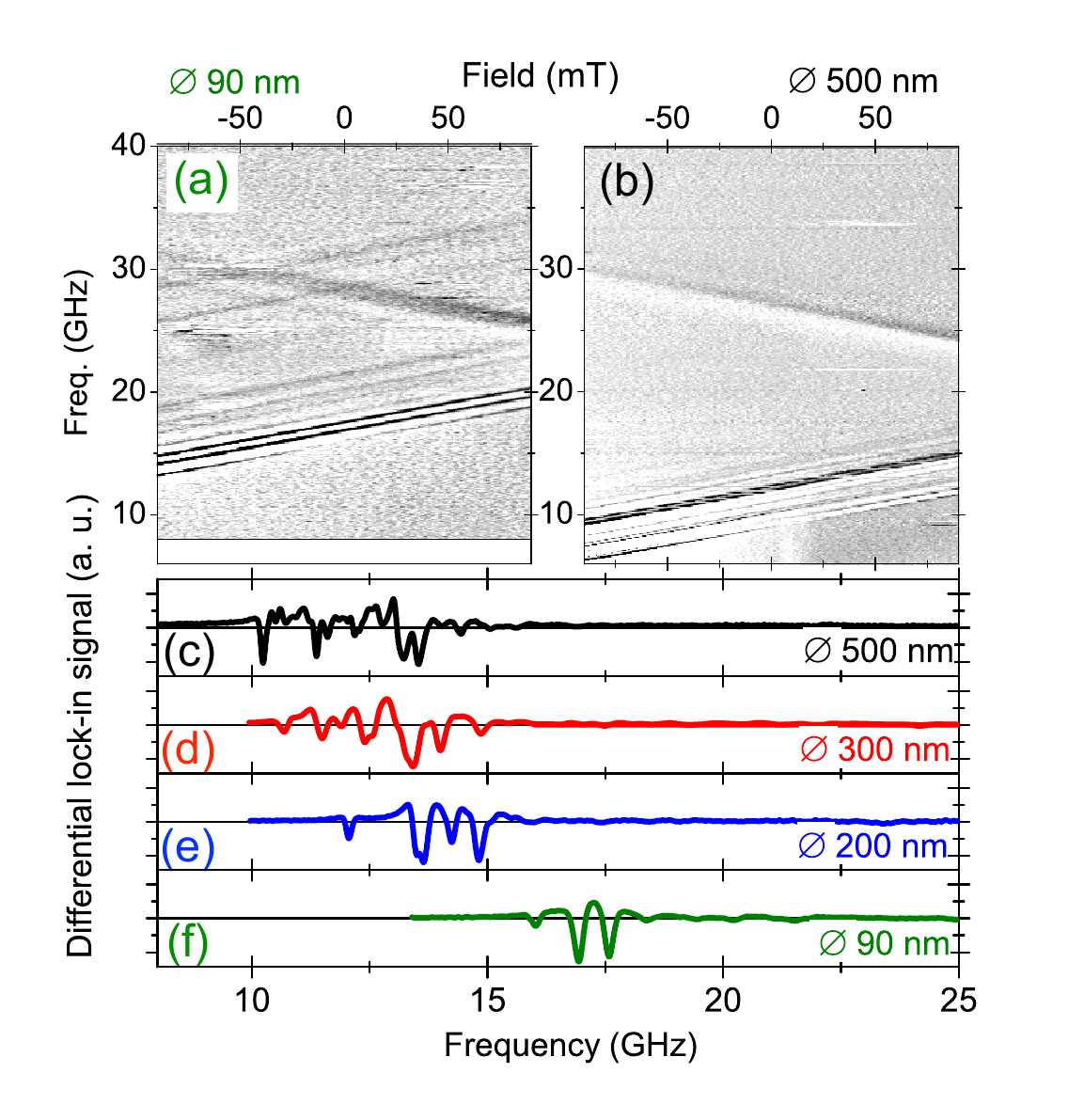}
\caption{(Color online). Dependence of the field derivative of the \textit{ac} voltage over the device diameter. Top panels: full spectral dependence in the window $[6,~40 ~\textrm{GHz}]\times [-90,~90~\textrm{mT}]$ for 90 nm (left) and 500 nm (right) diameter devices. Bottom panels: frequency dependence of the field derivative of the \textit{ac} voltage after self-conformal averaging for various device sizes. The signal amplitudes have been normalized to ease their comparison.}
\label{size}
\end{figure}

Conversely, the frequencies of the modes of the free layer are strongly affected by the device diameter (Fig.~\ref{size}). First, the modes are pushed to higher frequencies as the device is shrunk. At remanence, the lowest frequency mode is at $f_\textrm{QUP}=12.3~\textrm{GHz}$ for a diameter of 500 nm; it reaches 19.5 GHz for 60 nm devices (not shown). Second, the frequency spacing between the free layer modes increases substantially when downsizing the device.  

The first effect -- increase of $f_\textrm{QUP}$ at downscaling -- is indicative of a dependence of some effective fields with the device diameter. Among the effective fields, the only ones that vary with the diameter are the exchange fields \textcolor{black}{(positive contribution to the frequency $f_\textrm{QUP}$ if magnetization is non-uniform)}, the demagnetizing fields \textcolor{black}{(positive contribution to the frequency $f_\textrm{QUP}$ at downscaling)} and \textcolor{black}{the local effective anisotropy fields in case some process damages alter locally the interface anisotropy at the perimeter of the free layer (negative contribution to the frequency $f_\textrm{QUP}$ at downscaling) or alter the local magnetization of the rim of the free layer (positive contribution to the frequency $f_\textrm{QUP}$ at downscaling). The exchange fields are related to the non uniformities of either the static configuration --the fact that the AP state is not perfectly uniform as inferred previously from the loop in Fig.~\ref{T}-- or non uniformities of the dynamic magnetization --i.e. the fact that the quasi-uniform mode is not a strictly uniform mode--}. If the frequency increase was due to the sole demagnetizing effects, it could be estimated from the demagnetizing factors of disks \cite{mizunuma_size_2013} which are $N_z \approx 1- (3 \pi / 8) t /a$, where $t$ and $a$ are the thickness and radius of the free layer. However a $f_\textrm{QUP}$ against $1/a$ plot (not shown) has a perceivable curvature near all sizes; an unwise linear fit through $f_\textrm{QUP}$ against the expected $\gamma N_z$ would give a slope of $\approx2.5 ~\textrm{T}$, which is obviously too large for the magnetization of the free layer. \textcolor{black}{This indicates that the sole change of the global shape anisotropy with the device diameter is insufficient to account for the increase of $f_\textrm{QUP}$ at downscaling:  exchange contributions or non uniformities induced by process damages also contribute to the frequencies. Exchange contributions should not contribute for the largest devices, however even for those devices the experimental frequencies are larger than the ones expected from global shape anisotropy only, which argues for some process damages. Since the TMR is almost independent of the device size \cite{devolder_size_2016}, we can reasonably assume that the MgO/FeCoB interface is not substantially affected by the patterning and that consequently the interface anisotropy is essentially preserved at the rim of the free layer.}  \textcolor{black}{We conclude that part of the increase of $f_\textrm{QUP}$ at downscaling is due to a reduced magnetization (magnetically "dead" or weak zone) near the edges of the free layer. This interpretation is probably very much stack and process technology dependent, hence it should not be considered as general.}

The second effect -- the increased frequency spacing between the modes at small diameters -- is the expected effect of the confinement of the spin waves and the resulting increase of the exchange contribution to the mode frequencies\cite{devolder_exchange_2016}. The eigenmodes of perpendicularly magnetized circular disks are well understood and can be described analytically in a semi-quantitative manner \cite{kakazei_spin-wave_2004, klein_ferromagnetic_2008, arias_theory_2009, naletov_identification_2011, nedukh_standing_2013, munira_calculation_2015}. The frequency spacing between the lowest frequency modes scales with $\gamma_0 H_J$ where $H_J=\frac{2 A k^2}{\mu_0 M_S}$ is a generalized exchange field with $A$ the exchange stiffness. The effective wavevector $k$ is reminiscent of the lateral confinement and reads $k^2= ({u_2^2-u_1^2}) /a^2 \approx 9 /a^2 $ where $u_1$ and $u_2$ are the first zeros of the first and second  Bessel functions \cite{klein_ferromagnetic_2008}. The lowest frequency spinwave modes can be resolved only if their frequency spacing is comparable or greater than their linewidth $2 \alpha \gamma_0(H_z+H_k-M_s+H_J)$ (see appendix).

This condition can be used to define a critical device diameter:
\begin{equation} a_\textrm{crit}^2  = \frac{9 A}{ \alpha \mu_0 M_S (H_z+ H_k-M_s)} 
\label{ModeSeparation}\end{equation}
For large devices with $a \gg a_\textrm{crit}$ we expect to observe a quasi-continuum of overlapping modes above $f_\textrm{QUP}$, while discrete non-overlapping modes are anticipated in the opposite limit. Typical parameters of an FeCoB-based free layer include a magnetization of $\mu_0 M_s=1.2$ T and a Gilbert damping of\cite{devolder_time-resolved_2016} $\alpha = 0.01$. From the quasi-uniform mode frequency, we can get our effective anisotropy which is $H_k-M_s = 330$ kA/m. 
If the exchange stiffness of the free layer was bulk-like (i.e. $A=22$ pJ/m) like in ref.~\onlinecite{devolder_exchange_2016}, the critical diameter would be $2a_\textrm{crit} = 444~\textrm{nm}$. In practice the small frequency spacings between the modes of our samples indicates that the exchange stiffness of our free layer is in the range of 6-7 pJ/m i.e. well below the bulk value. \textcolor{black} {This estimate of the exchange stiffness was deduced assuming perfectly pinned boundary conditions for the spin-waves at the device edge, which is a questionable \cite{guslienko_effective_2002} assumption.} However the exchange stiffness is anyway weak in the free layer and this can be also qualitatively seen directly from the spin wave spectroscopy: indeed the frequency spacing of the lowest modes of the reference layer system is typically twice larger that the frequency spacing of the lowest modes of the free layer [see for instance Fig.~3(b)].
\textcolor{black}{While the reason for this small value of the free layer exchange stiffness is not entirely clear, we emphasize that having such a small exchange stiffness is not uncommon in magnetic systems that comprise only a small number of atomic layers, starting for instance from 2 pJ/m for a single layer of iron~\cite{romming_field-dependent_2015}}. Anyway with these parameters, we expect a clear separation of the lowest frequency modes at remanence provided that the device diameter is much smaller than $2a_\textrm{crit} = 250~\textrm{nm}$. 

In practice for 300 and 500 nm devices a fine structure can still be detected in the spin-wave density of states [see Fig.~\ref{size}(c)] but it is hard to count the modes and guess their frequencies out of this fine structure. In the remainder of this paper, we shall thus only consider devices of diameter less than 200 nm, in which the different spin-wave modes can be unambiguously resolved [see Fig.~\ref{size}(e-f)].

%

\section{Lineshape evolutions with bias and extraction of the Gilbert damping}
Let us now compare the shapes of the experimental rectified signal with those expected (see appendix). For that purpose we harness the different bias dependences (Table~\ref{PMAmacrospinLINESHAPES}) of the rectified signals $V_{1,~ac}$ and $V_{2,~ac}$ to isolate each of them in the experimental signal $V_{ac}$. We identify $V_{1,~ac}$ to the experimental curve $V_{ac}$ measured at ${V_{dc}=0}$, and we construct an estimate of $V_{2,~ac}$ by subtracting $V_{ac}$ measured at ${V_{dc}=0}$ from that measured at ${V_{dc} \neq 0}$ (see Table~\ref{PMAmacrospinLINESHAPES}). \textcolor{black}{Note because of this subtraction, any dependence of the spin-wave frequency with the \textit{dc} voltage will prevent the measurement of the voltage dependence of the damping factor or of the voltage dependence of the exciting torques.}

\begin{figure}
\includegraphics[width=9 cm]{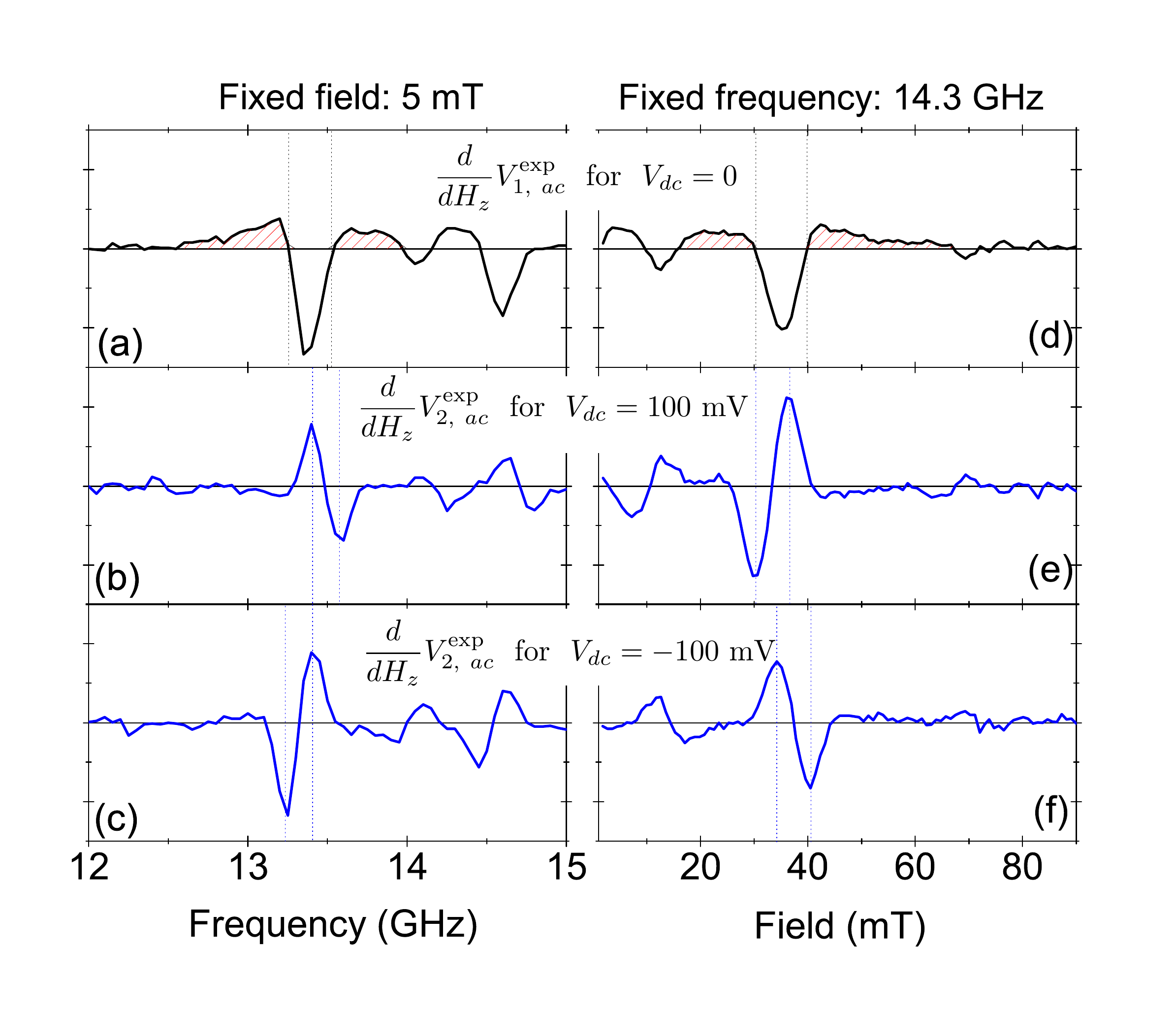}
\caption{(Color online). Rectified \textit{ac} signals versus frequency at fixed applied field (left panels) or versus field at fixed frequency (right panels). The plotted data are $\frac{d}{dH_z} V_{1,~ac}$ (black, panels a and d) 
and  $\frac{d}{dH_z} V_{2,~ac}$ (blue, panels b, c, e and f) as estimated according to the formulas of Table \ref{PMAmacrospinLINESHAPES}. 
The arbitrary vertical scale is the same for all panels. The dotted black lines are separated by 9.5 mT and 270 MHz. The  dotted blue lines are separated by 6.1 mT or 170 MHz. These dotted lines correspond to the expected linewidth for a damping of 0.011. The panel (b) is measured for a device different from that of the other panels.}
\label{BiasDep}
\end{figure}

We illustrate this procedure in Fig.~\ref{BiasDep} in which we plot the field derivatives of the so-calculated $V_{1,~ac}$ and $V_{2,~ac}$ rectified signals in both fixed field or fixed frequency experimental conditions for a device of diameter 90 nm. We center the curves on the second lowest frequency eigenmode since it provides the largest signals and it is reasonably separated from both the quasi-uniform precession mode and from the other high order modes. The obtained experimental $V_{1,~ac}$ curves [see Fig.~\ref{BiasDep} (a) and (d)] have the expected line shapes (see appendix) with a negative peak surrounded by two tiny positive halos (areas shaded in red). The separation between the two zero crossings is $9.5 \pm 0.5~\textrm{mT}$ or $285\pm 10~\textrm{MHz}$.
The obtained experimental $V_{2,~ac}$ curves also have the expected line shape of the derivative of a Lorentzian distribution (see appendix). As expected, the sign of the response changes with the \textit{dc} bias voltage. The separation between the positive and negative maxima of the distribution are $6.1\pm 0.5~\textrm{mT}$ and $170\pm 10~\textrm{MHz}$. 

These four different ways of measuring the linewidths [Fig.~\ref{BiasDep} (a, d, b, e)] are consistent with a free layer damping of $\alpha = 0.011\pm 0.0003$. Indeed this value of damping would predict linewidths of $2\alpha  f = 295~\textrm{MHz}$ and  $2 \mu_0 \alpha  \omega / \gamma_0 =9.54~ \textrm{mT}$ (materialized as black bars in [Fig.~\ref{BiasDep} (a, d) and (d)]) and $1.15 \alpha  f= 171~\textrm{MHz}$ or  $1.15\alpha  \omega / \gamma_0 = 6.21~\textrm{mT}$ (materialized as blue bars in [Fig.~\ref{BiasDep} (b, c, e, f)]).

\begin{figure}
\includegraphics[width=9 cm]{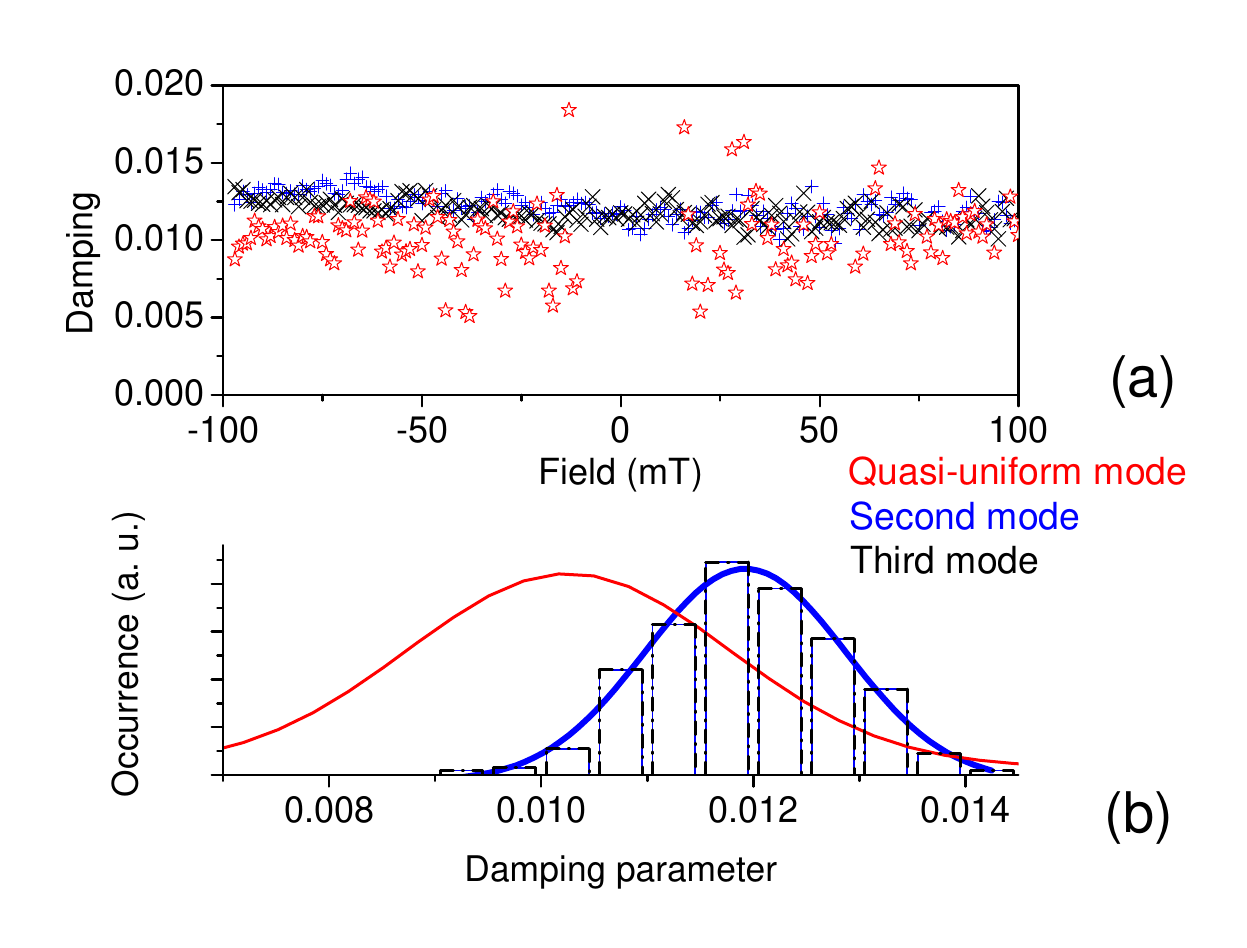}
\caption{\textcolor{black}{(Color online). Statistical view of the Gilbert damping parameters obtained from the fitting of the three lowest spin-wave resonances of a device of 90 nm diameter. For each applied field, the spectral line shapes of the quasi-uniform mode is fitted using $\frac{d}{dH_z} V_{2,~ac}$ function (red symbols) while the lineshapes of the two next modes are fitted with  $\frac{d}{dH_z} V_{1,~ac}$ function (blue and black). The panel (a) gathers the values of the damping parameters that best fit each spectrum recorded at a given applied field. Panel (b) displays the histogram of the distribution of these estimates of the Gilbert damping for the non-uniform modes (dashed-dotted line histogram and its blue Gaussian guide to the eye, of half width 0.0009) and a Gaussian fit (red curve) of the distribution for the quasi-uniform mode, of half width 0.0015.}}
\label{DampingVersusMode}
\end{figure}

This proposed value of damping is also consistent with the linewidths of higher order spin waves that appear at larger frequencies but with a lower signal. 
\textcolor{black} {This is illustrated in fig.~\ref{DampingVersusMode} where a comparison is drawn between the values of the damping estimated for each applied field from the second (and most intense) mode and from the third mode for a device of diameter 80 nm. The used procedure is a direct fit of the experimental lineshapes to the derivative of Eq.~\ref{ReChi} with the damping, the resonance frequency and the signal amplitude as free parameters. For this specific device, the estimates of the damping parameter are subjected to a random error of standard deviation 0.0018 around a mean value of $\alpha=0.0119$. \textcolor{black}{It is interesting to note that the non-local contributions \cite{nembach_mode-_2013, wang_phenomenological_2015} to the damping expected for the relatively large wavevectors of the second and third spin-wave modes seem to be too small to be observed in our samples}. Note that as mentioned earlier, the same procedure can in principle not be applied to the quasi-uniform mode since it exhibits a strong dependence of the mode amplitude with the field which invalidates the procedure to some extent. For the sake of completeness of this paper, we have anyway fitted the experimental QUP lineshapes with the field derivative of  Eq.~\ref{ImChi}; this is not possible near zero field, as the corresponding signal vanishes. The value of the Gilbert damping that would be illegitimately deduced would be $0.01$, i.e. 20\% lower than the correct value. Besides, the estimates from the QUP mode would exhibit a substantially larger spread in the fit results [compare the histograms in in fig.~\ref{DampingVersusMode}(b)]. For these two reasons, we consider that the reliable estimate of the damping is the one extracted from the non-uniform modes.}

Above 100 mV of \textit{dc} bias, the amplitude of the constructed experimental $V_{2,~ac}$ start to depart from proportionality with $V_{dc}$ and a frequency shift is observed, as expected when \textit{dc} field-like spin torques are applied. This comes with by a distortion of the line shape, probably linked to the modification of the spin waves lifetimes by spin-transfer torque as commonly observed in in-plane magnetized MTJs \cite{petit_spin-torque_2007}.

\section{Summary and conclusions} 
In this paper, we have studied how to use \textit{rf}-voltage-induced ferromagnetic resonance to study the spin-wave density of states and the Gilbert damping in perpendicularly magnetized disks embodied in magnetic tunnel junctions. We have applied the field along the easy axis to preserve the cylindrical symmetry of the magnetization energy functional. The interest of this configuration is that all the current-induced torques that potentially excite the dynamics yield the same type of susceptibility spectral shape. Additionally, this configuration is the sole in which the applied field and the frequency play similar roles near FMR so that consistency crosschecks between variable-field and variable-frequency experiments can be performed to reveal and suppress potential experimental artefacts. 

Working in a situation in which the fixed layer and the free layer are oppositely magnetized in a convenient way to classify the spin-waves according to their hosting layer, as the two sub-systems have opposite eigenmode frequency-versus-field slopes. The \textit{dc} bias dependence of the signal of the quasi-uniform mode is peculiar and can be used to ambiguously identify the free layer quasi-uniform mode in the manifold of spin-waves. The non-uniform (higher order) spin-waves are easier to analyze, as  their amplitudes weakly depend on the applied field so that field differentiation can be used safely for background subtraction. Optionally, the dynamic range of the experiment can be improved by self-conformal averaging of the resonance spectra.

The unambiguous identification of the spin-wave frequencies requires devices that are sufficiently small to avoid that the spinwave modes overlap into a quasi-continuous density of states. The critical device size is set by the exchange stiffness, the damping, the magnetization and the effective anisotropy field. In practice, device diameters below 200 nm are needed in our low-damped FeCoB-based PMA system. 

For each spin-wave mode, the \textit{rf}-voltage-induced spin-wave spectra contain contributions from two different physical mechanisms. The first one is the standard STT-FMR-like signal, whose spectral shape is a linear transverse susceptibility term. It is independent from the \textit{dc} voltage applied across the MTJ. The second one is a variation of the time-averaged magnetic configuration when the \textit{rf} voltage is applied. It is proportional to the \textit{dc} voltage applied across the MTJ and it has the spectral shape of a non-linear longitudinal susceptibility. The bias dependence can be used to separate these two signals. The analysis of their spectral shape yields the Gilbert damping within the precessing layer. \textcolor{black}{A single value of the damping factor is found to account for the lineshapes of all studied spin-waves.}

The spectra of \textit{rf}-voltage-induced rectified voltages for a vanishing \textit{dc} voltage bias are in principle sufficient to get the Gilbert damping of the dynamically active layer. However as microwave methods are prone to artefacts, a consistency check exploiting the bias dependence of the resonance spectra is useful for a consolidation of the numerical estimation of the damping.

This work was supported in part by the Samsung Global MRAM Innovation Program, who provided also the samples. Critical discussions with Vladimir Nikitin, Jean-Paul Adam, Joo-Von Kim and Paul Bouquin are acknowledged.

%

%

\section{Appendix: susceptibilities in an idealized PMA film}
In this appendix, our aim is to determine the transverse and longitudinal microwave susceptibility versus frequency $f$ and static field $H_z$ for a PMA film as a response to an harmonic transverse field $h_x \cos(\omega t ) \vec{e_{x}}$. We shall write the equations with this transverse field  $h_x$ but any other effect that yields a torque possessing a component transverse to the static  magnetization will yield similar lineshapes. This includes current-induced Oersted-Ampere fields but also Slonczewski STT and field-like STT as soon as the reference layer magnetization $\vec{M}_\textrm{ref}$ is not strictly collinear with that of the free layer $\vec{M}_\textrm{free}$. The susceptibility tensor will be used to deduce the shape of the line expected in \textit{rf}-voltage-induced FMR, as summarized in Table~\ref{PMAmacrospinLINESHAPES}. Throughout this appendix, we assume a \textit{dc} field $H_z$ perfectly perpendicular to the plane and a free layer magnetization $\vec{M}=M_z \vec{e}_z+m_x \vec{e}_x + m_y \vec{e}_y$, where the transverse terms are assumed small and written as complex numbers in the frequency space. For conveniency, we will use the notation $H'=H_z + H_k -M_s$. We shall also write the frequencies in field units and define $\omega' = \omega/\gamma_0$. This is meant to emphasize the fact that the generalized field $H'$ and the generalized frequency $\omega'$ play very similar roles in the FMR of perpendicularly magnetized macrospin when near resonnance. We shall systematically assume that $\alpha \ll 1$ and only keep the lowest order of the damping terms in the equations.
In sections \ref{transverse} -\ref{lineshapes} we make the macrospin approximation. This approximation is discussed in section \ref{beyond}.

\subsection{Transverse linear susceptibility} \label{transverse}
Following the usual procedure, we project the linearized Landau-Lifshitz-Gilbert equation along $\vec{e}_{x}$ et $\vec{e}_{y}$ :
$$
\left\{\begin{array}{ccl}H'(m_y+\alpha m_x)+im_x\omega' &=& h_x M_s\alpha\\
H'(m_x-\alpha m_y)-im_y\omega' &=& h_x M_s\end{array}\right.
$$
We then invert this system of equations to get the susceptibilities $m_x=\chi_{xx}h_{x}$ and $m_y=\chi_{yx}h_{x}$. They are: 


 \begin{equation} \chi_{xx} = \frac{  {M_s} H' 
 } {{H'}^2 -\omega'^2 +2 \text{i} \alpha  H' \omega '  }  \label{chixx} \end{equation}
   
   and 
 \begin{equation} \chi_{yx} = - \text{i} \frac{ {M_s} \omega'}{ {H'}^2 -\omega'^2 +2 \text{i}   \alpha  H' \omega '} \label{chiyx}\end{equation}
 
Several points are worth to remind: \\
(i) The \textit{dc} transverse susceptibilities are $\chi_{xx}^{dc}= M_s / H'$ and $\chi_{yx}^{dc}= 0$. \\
(ii) The in-phase transverse susceptibility $\chi_{xx}$ is peaked at the FMR condition $\omega' = H'$. It reaches $\chi_{xx}^{FMR} = - \frac{1} {2 i \alpha} \chi_{xx}^{dc}$.
(iii) When near the FMR condition, we have $\chi_{yx} \approx - i \chi_{xx} \label{CircularPrecession}$. Hence the two transverse components of the magnetization are in quadrature and the forced precession is essentially circular. Note that this holds true despite the fact that the pumping field is linearly polarized (i.e. along $(x)$ only). It would also remain true for other (e.g STT) pumping torques. \\

From Eq.~\ref{chixx} we deduce the classical expressions for the real and imaginary parts of the transverse susceptibility:
 \begin{equation}  \Re e (\chi_{xx})=\frac{M_sH'({H'}^2-\omega'^2)}{4\alpha^2{H'}^2 \omega'^2+({H'}^2-\omega'^2)^2} \label {ReChi}  \end{equation}
 \begin{equation}  \Im m (\chi_{xx})= -  \frac{2 \alpha M_s\omega'{H'}^2}{4\alpha^2{H'}^2 \omega'^2+({H'}^2 -\omega'^2)^2}  \label {ImChi} \end{equation}
 
The lineshapes given by the above expressions are shown in Fig.~\ref{TypicalLineProfile}. Their main properties are summarized in Table~\ref{PMAmacrospinLINESHAPES}.

\subsection{Longitudinal non-linear susceptibility} \label{longitudinal}
Let us now express the non-linear change of the longitudinal magnetization $\Delta M_z$ that occurs due to the precession. This can be viewed as an \textit{rf}-induced reduction of the remanence. Using the circularity of the precession near the FMR resonance and the conservation of the magnetization norm to second order in $m_{x,~y}$, one gets:

 \begin{equation} \Delta M_z \approx \frac{|| h_{x} || ^2}{2M_S} \frac{M_s^2 H'^2}{[{H'}^2-\omega'^2]^2 + 4 {\alpha}^2 {H'}^2 \omega'^2} \label{DeltaMz}Ê\end{equation}
where $|| h_{x} ||$ is the (constant) amplitude of the applied \textit{rf} field.
It is worth noticing that the longitudinal loss of the magnetization is \textit{stationary (constant in time)} despite the fact that the magnetization precesses continuously. 

\subsection{Lineshapes and linewidths of the susceptibiliies and their derivatives} \label{lineshapes}
\begin{figure}
\includegraphics[width=9.0 cm]{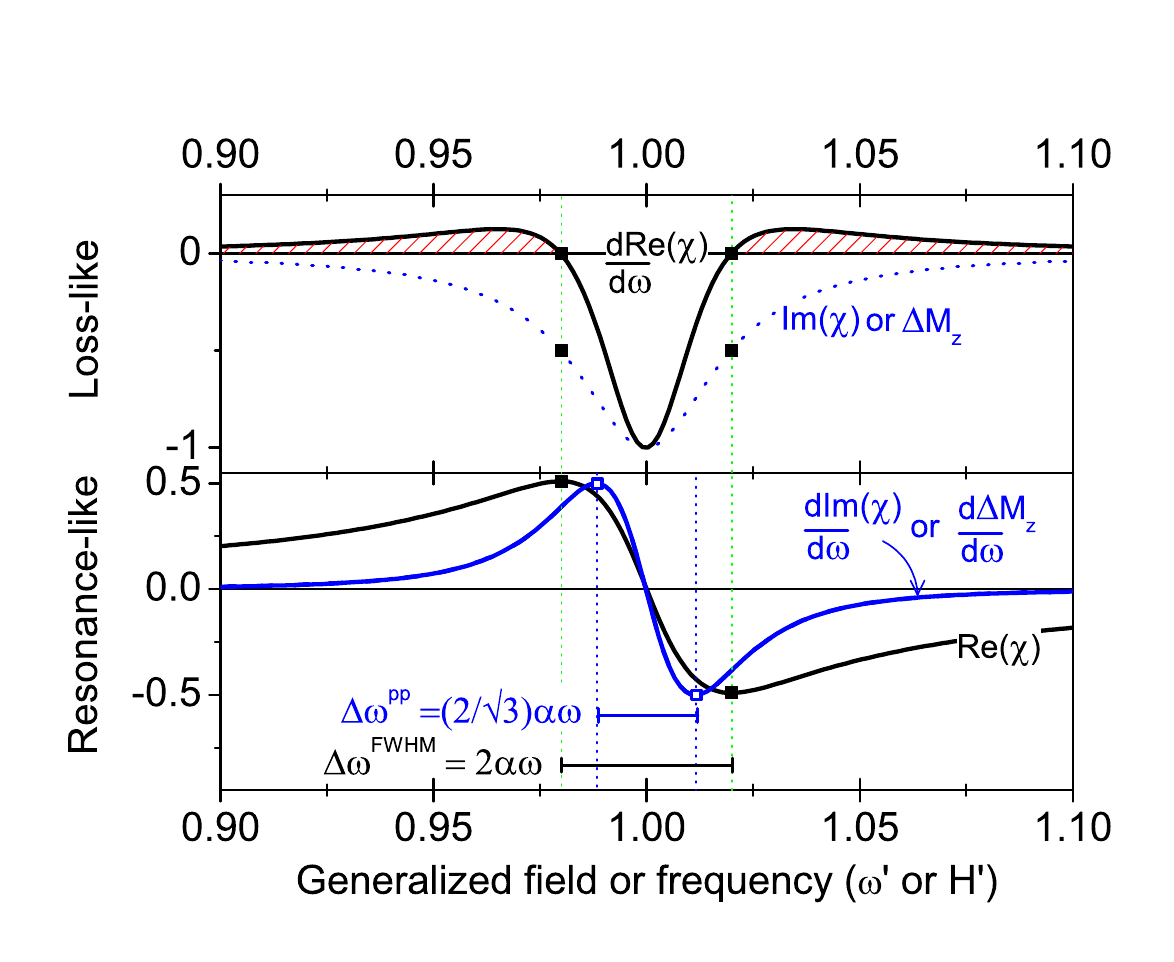}
\caption{(Color online). Transverse susceptibilities, longitudinal susceptibility and their derivatives in the PMA macrospin model. The curves are plotted for $\alpha=0.02$ and a resonance condition of unity. The responses amplitudes have been normalized to ease the comparison between the different lineshapes. The black horizontal segment in the bottom panel is the FWHM of $\Im m(\chi_{xx})$ and  $\Delta M_z$ or equivalently the peak-to-peak separation of $\Re e (\chi_{xx})$ (also sketched as the black square dots). The blue segment sketches the peak-to-peak separation of $\frac{d \Im m(\chi_{xx})}{d \omega'}$ (also sketched as the empty blue square dots). The area shaped in red is the positive halo that surrounds the main (negative) peak in $\frac{d \Re e(\chi_{xx})}{d \omega'}$. }
\label{TypicalLineProfile}
\end{figure}

The different susceptibility expressions are plotted in Fig.~\ref{TypicalLineProfile}. The lineshape of the functions $\Delta M_z$ and $\chi_{xx}$ are essentially determined by their denominator which are the fast varying functions of eqs.~\ref{ImChi} and \ref{DeltaMz}. As $\chi_{xx}$ and $\Delta M_z$ are two signatures of the same resonance process, their denominators are equal (see eqs.~\ref{ImChi} and \ref{DeltaMz}) and a simple algebra confirms that $\Im m(\chi_{xx})$ and $\Delta M_z$ lead to the same frequency or field linewidths (Half Width at Half Maximum) which are:
\begin{equation} 
\Delta \omega = 2 \alpha \omega \textrm{~~or~~equivalently,~~}  \Delta H_z = 2 \alpha H' \label{linewidth}
 \end{equation}
Note that $\Delta \omega$ (resp. $\Delta H_z$) is also the frequency (resp. field) spacing between the positive maximum and negative maximum of $\Re e(\chi_{xx})$ about the FMR condition.

We stress that this linewidth is \textit{different} from that obtained by conventional FMR in which people examine the \textit{derivative} of the absorption signal $\frac{d  \Im m (\chi_{xx})}{d H_z}$ versus $H_z$. The peak-to-peak separation of the conventional FMR signal is $\Delta H_z = \frac{2}{\sqrt{3}} \alpha H' = \frac{2}{\sqrt{3}} \alpha \omega' $. The factor $2 / \sqrt{3}$ is 1.1547. 
The spectral shapes of $\Re e(\chi_{xx})$  and  $\Delta M_z$ as deduced above are to be used in the main part of the paper to describe respectively $V_{1,~ac}$ and $V_{2,~ac}$ (Table~\ref{PMAmacrospinLINESHAPES}).


%

\subsection{Linewidth beyond the macrospin approximation} \label{beyond}
Some words of caution are needed as the calculations done so far the appendix are for the uniform precession mode of a macrospin, while most of our experimental results were obtained on the higher order (non-uniform) spin-waves. When modeling non-uniform spin-waves, one needs to take into account additional exchange and dipole-dipole terms.

For non-uniform spin-waves in perpendicularly magnetized system, the exchange fields related to the non uniformity of the dynamical magnetizations $m_x$ and $m_y$ can be added to $H'$ to form a new generalized field $\tilde{H} = H + H_k - M_s + \frac{2A k^2}{\mu_0 M_S}$, where $k$ is a generalized wavevector\cite{klein_ferromagnetic_2008, naletov_identification_2011, munira_calculation_2015}. The additional exchange contributions act on $m_x$ and $m_y$ on equal footing, hence they maintain the circularity of the precession. The Gilbert linewidth for a non-uniform spin-wave is simply\cite{slavin_approximate_2005, kim_stochastic_2006}: 

\begin{equation}
\Delta \omega = \alpha \omega_k \frac{\partial \omega_k}{\gamma_0  \partial \tilde{H}} \label{circularLinewidth}
\end{equation}
Where this equation holds as $\tilde{H}$ is a circular term. As a result of Eq.~\ref{circularLinewidth} the proportionality of an eigemode linewidth to its frequency (Eq.~\ref{linewidth}) is not broken by the exchange contributions in non-uniform spin-waves.

Conversely, the directional nature of the dipole-dipole interaction is such that the related effective fields act differently on the two dynamical magnetizations $m_x$ and $m_y$ for spin-waves having a non-radial character. As a result the dynamic demagnetizing fields can not be simply added to the circular precession $H'$ term and they induce some ellipticity of the precession of the non-uniform spin-waves. Dipole-dipole interactions make the eigenmode frequency non-linear with the field (see Eq. 52 in ref.~\onlinecite{kalinikos_theory_1986}). For the lowest lying non-uniform spin-wave the dipole-dipole stiffness field is $M_S k t /2$ with $k \approx \pi/a$. It is negligible against the generalized field $\tilde{H}$ for the thickness used in practice in PMA systems meant for spin-torque applications. The precession stays therefore essentially circular for all the modes observed experimentally here; we will thus consider that it is legitimate to use Eq.~\ref{linewidth} and deduce the damping from the ratio of the half frequency linewidth to the eigenmode frequency. 

\textcolor{black}{Finally, we would like to mention that our method is not restricted to the PMA materials only: it should hold when the considered spin-waves are quasi-circular. In particular, this is the case of exchange-dominated spin waves in in-plane magnetized systems \cite{dvornik_dispersion_2011}.}


%

\end{document}